\documentclass[
aps,nofootinbib,showpacs,showkeys,preprint
] {revtex4}

\usepackage{epsf,epsfig,subfigure,graphicx,amsmath,amssymb}
\usepackage{color}

 \def\LN{LN$\chi$}

\begin{document}

\title{\Large\bf Two dark matter components in
dark matter extension of the minimal supersymmetric standard model and the high energy\\
positron spectrum in PAMELA/HEAT data }

\author{Ji-Haeng Huh\email{jhhuh@phya.snu.ac.kr}, Jihn E. Kim\email{jekim@ctp.snu.ac.kr} and Bumseok Kyae}
\affiliation{ Department of Physics and Astronomy and Center for Theoretical Physics, Seoul National University, Seoul 151-747, Korea
 }
\begin{abstract}
We present the dark matter (DM) extension (by $N$) of the minimal supersymmetric standard model to give the recent trend of the high energy positron spectrum of the PAMELA/HEAT experiments. If the trend survives by future experiments, the MSSM needs to be extended. Here, we minimally extend the MSSM with one more DM component  $N$ together with a heavy lepton $E$, and introduce the coupling $e_R E_R^c N_R$. This coupling naturally appears in the flipped SU(5) GUT models. The model contains the discrete symmetry $Z_6$, and for some parameter ranges there result two DM components. For the MSSM fields, the conventional $R$-parity, which is a subgroup of $Z_6$,
is preserved. We also present the needed parameter ranges of these additional particles.
\end{abstract}

\pacs{95.35.+d, 12.60.Jv, 14.80.-j, 95.30.Cq}

\keywords{High energy galactic positrons, Two DM components, PAMELA data}
\maketitle


The existence of dark matter (DM) at the 23\% level of the closure density \cite{WMAP08} is largely accepted by the observations of the flat rotation curves of the velocities of halo stars, the simulation of the bullet cluster collision, and gravitational lensing experiments. So, the identification of the cosmological DM is of the prime importance in particle physics and cosmology. If indeed the DM particles of $O(100)$ GeV mass with the weak interaction strength are abundant in galaxies, high energy positrons, antiprotons and gamma rays from DM annihilation has been predicted for a long time \cite{highEpartDM}. If the DM annihilation is confirmed, the {\it very weakly} interacting DM possibility \cite{CKKR} is ruled out \cite{CKLS}.

The satellite PAMELA experiment has already started to probe an interesting SUSY parameter space. Their recent report on the high energy positron observation, above 10 GeV up to 60 GeV \cite{PAMELAexp}, has already spurred a great deal of attention \cite{Bergstrom08,Barger08}. In fact, the same trend has been noticed earlier in the balloon-borne HEAT experiment but with larger error bars \cite{HEAT95}. The charactersistic of the PAMELA/HEAT data between 10--50 GeV is a slightly rising positron flux, $e^+/(e^++e^-)\sim O(0.1)$. If confirmed by independent observations such as PEBS baloon experiments \cite{PEBS} and the AMS-02 experiments \cite{AMS02}, the implication is tantamount in that the most popular \LN\ DM scenario of  the minimal supersymmetric standard model (MSSM) maybe in jeopardy. Even though the possible astrophysical explanations have been presented in \cite{Pulsar}, here we focus on the particle physics explanation. On the other hand, we note that the same PAMELA data does not show any significant anti-proton excess \cite{PAMELAexp}.

\begin{figure}[!h]
\resizebox{1\columnwidth}{!}
{\includegraphics{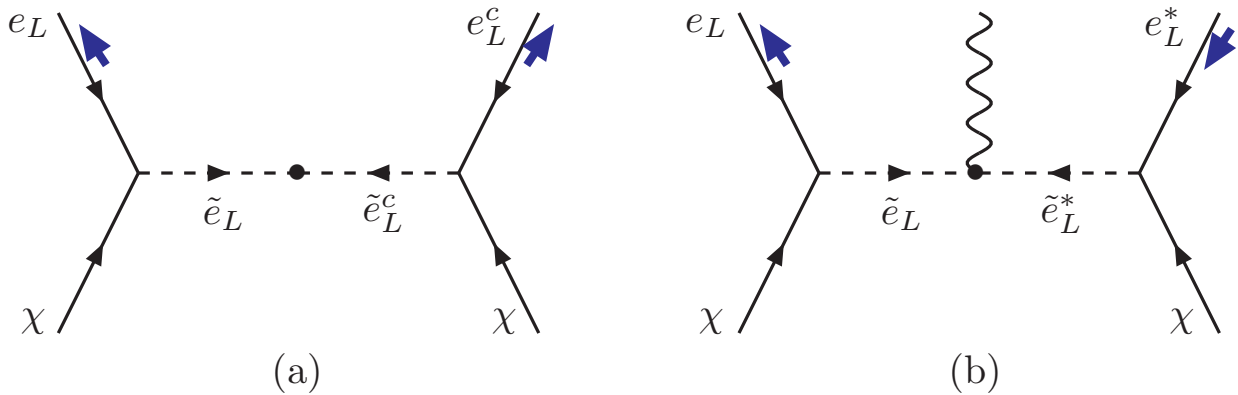}}
\caption{\it The bino-like neutralino annihilation. (a) The bullet carries an $SU(2)_W$ quantum number. (b) Here, the bullet can carry angular momentum 1. The helicities of electrons and positrons are shown by thick arrow lines.}\label{fig:oneneutral}
\end{figure}

The lightest neutralino $\chi$ (\LN) of the MSSM  has been favored as the cold dark matter (CDM) candidate. If the CDM is composed of just one component Majorana fermion such as the neutralino LSP, due to the Fermi statistics two annihilating LSPs of Fig. \ref{fig:oneneutral}(a), being the identical fermions, can come close if their spins are anti-parallel so that the angular momentum is zero, or in the $s$-wave state. Accordingly, when the neutralinos are annihilated to a fermion--antifermion $(f\bar f)$ pair, the spins of $f$ and $\bar f$ should be anti-parallel.
For the helicity flipping, the amplitude should necessarily acquire a factor $m_f$ in the $s$-wave state. Thus, co-annihilation of neutralinos into $c, b$, and $t$ quark pairs, if it is open, would dominate over all other channels in the MSSM. In view of the conventional MSSM CDM scenario, therefore, the recent reported PAMELA data on the high energy positron excesses is quite embarrassing because of the above difficulty faced in the MSSM CDM scenario.

This leads us to consider a minimal extension of the MSSM so as to keep its most desirable property {\it `supersymmetry'}. In the framework of spin-$\frac12$ CDMs, therefore, we extend the MSSM minimally to include two CDM components.

In the MSSM, if the \LN\ is bino denoted as $\chi$, then the cosmologically favored bino density in the universe is possible in the co-annihilation region \cite{BinoLSP}. In this paper we treat the \LN\ as the bino just for the sake of a concrete discussion. If the \LN\ contains the bino as a fraction, the discussion should be modified accordingly.

As commented above, two annihilating neutralinos of Fig. \ref{fig:oneneutral}(a), being the identical fermions, can come close if their spins are anti-parallel so that the angular momentum is zero. If the final electron and positron, going out back to back, have the same helicity, then they can make up the angular momentum zero. One such possibility shown in Fig. \ref{fig:oneneutral}(a) shows that it is highly suppressed because the bullet of Fig. \ref{fig:oneneutral}(a) carrying an $SU(2)_W$ quantum number $f_e\langle H_d^0\rangle m_{3/2}$ a high suppression factor. On the other hand, if the outgoing electron and positron carry the opposite helicities, their spin is one and by emitting a photon the three particles final state can make up angular momentum zero. However, in this case there is a coupling suppression of order $\alpha_{\rm em}/\pi$. This possibility of high energy positron plus photon has been suggested in Ref. \cite{Bergstrom08} where  a large enhancement factor of order $10^4$ is needed.  A scalar DM such as sneutrino \LN\ will have the same fate as the bino \LN\ in this regard.

One may consider a very heavy neutralino dominantly producing $W^+W^-$ so that subsequent $W$ decays provide the positrons. This requires a very large \LN\ mass O(10 TeV) which belongs to the neutral component of weak isospin $I=5$ multiplet to satisfy several constraints as a DM candidate \cite{cirelli}. This is also not very minimal in the sense that many component fields are introduced and SUSY particle masses are raised substantially. In addition, the intermediate $W^+W^-$ will give some excess anti-protons also.

This consideration presents a useful direction for constructing DM models with the rising high energy positron spectrum above 10 GeV. Obviously, a Dirac particle DM can be considered \cite{Gondolo08}. Also a spin-1 DM \cite{Barger08} can overcome the bino-like \LN\ difficulty, but here the spin-1 DM \cite{Barger08} would give some excess anti-protons also.

Thus, we extend the MSSM so that the DM annihilation produces high energy positrons but not excess anti-protons, thus providing a possible explanation of the PAMELA/HEAT data. This needs a special treatment of color singlet particles compared to quarks. So let us consider the next MSSM \cite{NMSSM} extended by a neutral singlet $N$ ($N_{\rm DM}$MSSM). $N$ is split into two Weyl spinors $\{N_R, N_L\}$. We also introduce an $SU(2)$ {\it singlet} charged lepton $E$ ($=\{E_R,E_L\})$. Without introducing this singlet lepton, we cannot achieve the goal of producing excess positrons: This charged lepton makes $N$ stable in some region of the
parameter space as we will discuss below.

This $N_{\rm DM}$MSSM seems to be very simple in the sense that new particles needed beyond the MSSM are {\it minimal}, just $N$ and $E$. To explain the correct order of DM density and the production cross section of positrons, the weak interaction nature of WIMP must be kept. Let us introduce the continuous $R$-symmetry
so that the additional DM component $N$ remains light down to low energies. The weak hypercharge $Y$ and $R$ charge of the singlet fields are
\begin{equation}
\begin{array}{cccccc}
{\rm Superfields:}&\quad e_R&\quad N_R&\quad N^c_R&\quad E_R&\quad E_R^c \\
Y~ :&\quad -1&\quad 0&\quad 0&\quad -1&\quad +1 \\
R~ :&\quad +1&\quad \frac23&\quad -\frac23&\quad -\frac13&\quad
+\frac13
\end{array}\label{ENCharges}
\end{equation}
The $R$ charges of the MSSM fields are as usual: the quark and lepton superfields carry 1 and the Higgs superfields carry 0. The $R$ symmetry allows the superpotential,
\begin{equation}
W=fe_R E_R^c N_R + hN_R^3\label{eENfromR}
\end{equation}
where $f$ and $h$ are coupling constants, but the mass terms of $N$ and $E$ cannot be present in $W$. However, via the Giudice-Masiero mechanism \cite{Giudice88}, supergravity effects can generate the fermion masses if the $F$ term of a singlet $S$ is developed,
\begin{equation}
\int d^4\theta \frac{S^*}{M_P}\left(\lambda
E_RE_R^c+\lambda'N_RN^c_R+{\rm h.c.}\right).
\end{equation}
Similarly the MSSM $\mu$-term is also generated. These masses are assumed to be of order the gravitino mass $m_{3/2}$. $N_R$ does not develop a VEV. As usual, with the separate lepton number conservation the process $\mu\to e\gamma$ is forbidden. Without the separate lepton number conservation, the $U(1)_R$ allows $E_R^c$ couplings to $\mu_R$ and $\tau_R$ by couplings $f'$ and $f''$, in which case we need $|f'|\le 10^{-4}$. The $f''$ bound is much weaker. 

Not introducing $N$ couplings to quarks, we will not introduce excess anti-protons. From string compactification with the doublet-triplet splitting \cite{flipstring}, the flipped $SU(5)$ is best suited for this purpose \cite{flipWD}.

The mass terms induced by SUSY breaking can be written as an effective superpotential,
\begin{eqnarray}
W\supset m_{3/2} E_RE_R^c+m_{3/2}'N_RN_R^c
,\label{massmixing}
\end{eqnarray}
and soft SUSY breaking $A$ and $B$ terms are
\begin{eqnarray}
{\cal L}&\supset& m_{3/2}[a_f\tilde{e}_R \tilde{E}_R^c \tilde{N}_R+a_h\tilde{N}_R^3]\nonumber\\
&&
+m_{3/2}^2[b\tilde{E}_R\tilde{E}_R^c
+b'\tilde{N}_R\tilde{N}_R^c]
+{\rm h.c.},\label{ABterms}
\end{eqnarray} where $a$ and $b$ denote dimensionless couplings. Eq. (\ref{ABterms}) violates the continuous $R$ symmetry, leading to a discrete $Z_6$ symmetry. The $R$-parity ($Z_2$) is a part of this $Z_6$ symmetry. For the fields in Eq. (\ref{ENCharges}), the $(R{\rm
-parity},Z_6)$ charges are
\begin{eqnarray}
&\tilde e_R(-,3) ,~\tilde N_R(+,2),~\tilde{N}_R^c(+,4), ~\tilde
E_R(-,5),~\tilde E_R^c(-,1),~~~
\nonumber\\
&e_R(+,0) ,~N_R(-,5) ,~N_R^c(-,1),~E_R(+,2),~E_R^c(+,4) .~~~
\end{eqnarray}
The SM fields are given $Z_6=0$ and their superpartners are given $Z_6=3$, which is just the $R$-parity for this MSSM subset fields in our $N_{\rm DM}$MSSM. The proton longevity is protected by the $U(1)_R$. The $U(1)_R$ forbids $u_Rd_Rd_R$ and the dimension 5 operators $q_R^cq_R^cq_R^cl_R^c$ and $u_Ru_Rd_Re_R$ which are allowed by the conventional $R$-parity alone. If SUSY breaking induce them in the $N_{\rm DM}$MSSM, they must be highly suppressed since there are no simple diagrams for them. Because of the $R$-parity of $N$,  $N$ cannot be a candidate for the singlet heavy neutrino of the seesaw mechanism.

We intend to introduce two stable particles, one the \LN\ and the other the lightest $Z_6$ matter particle (LMP). For the superfields $\tilde N_R+N_R\theta$ and $\tilde E_R+E_R\theta$, we assume $M_{\tilde N}>m_N$, $M_{\tilde E}>m_E$, and $m_E>m_N$. We assume that $N$ is lighter than $\tilde e_R$ and $\tilde E_R$, and hence $N$ is taken as the LMP. The LMP $N$ carries $Z_6=5$ which cannot be made by SM particles (carrying $Z_6=0$) alone. If $\chi$ is much heavier than $N$, then the decay, $\chi\to 3N^c+e^++e^-$ suppressed by $\tilde e_R$, $E_R^c$, and $\tilde N_R$ propagators, is possible.\footnote{Consider the Feynman diagram, $\chi\to E e^+ N^c ({\rm by~}\tilde e~{\rm propagator})\to (e^-\tilde N)({\rm by}~ E~ {\rm decay})+e^+ N^c \to e^-+N^cN^c+ e^+ N^c$ (by $\tilde N\to N^cN^c$ of Eq. (2)).} Below, we consider the case of $N$ being lighter than $\chi$.  The decay rate is estimated as,
$\Gamma\sim g'^2|f|^4|h|^2M_\chi^{11}/M^4_{\tilde N}M^4_{\tilde e}m^2_{E}$. But we require $M_\chi<3m_N$ for a successful two DM components in the universe. In this way, we have two DM components, $\chi$ and $N$.  Now, we take the bino as the \LN\ and the fermionic partner $N$ as the LMP.

\begin{figure}[!h]
\resizebox{0.6\columnwidth}{!} {\includegraphics{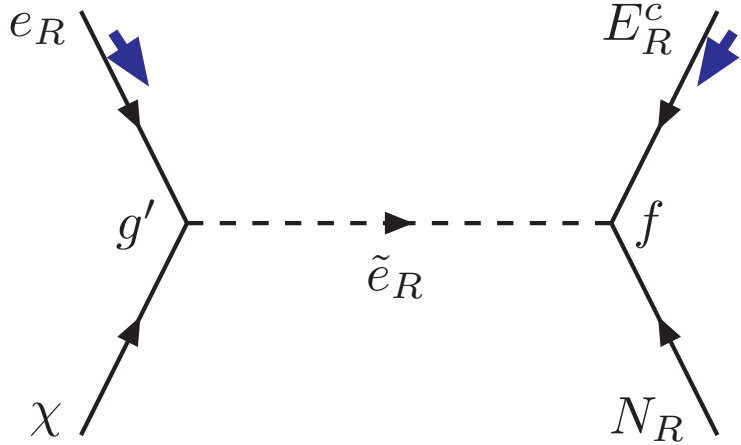}}
\caption{\it  A typical diagram for a bibo $\chi$ annihilation. $N_RN_R^c$ annihilation to $e^+e^-$ is also possible.}\label{fig:chiNRscatt}
\end{figure}
\begin{figure}[!h]
\resizebox{0.6\columnwidth}{!} {\includegraphics{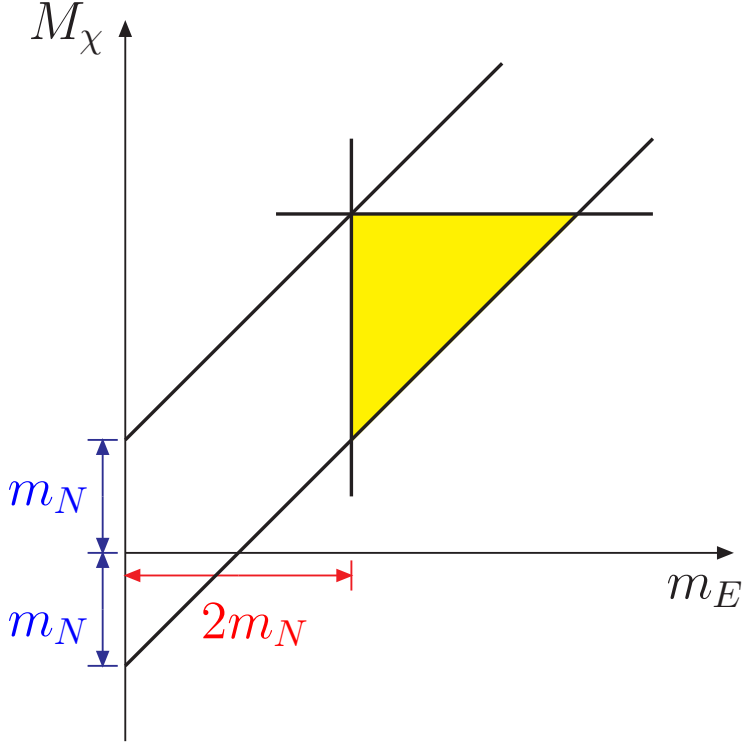}}
\caption{\it   In the $M_{\chi}-m_E$ plane, the kinematically allowed mass region is shaded for a typical mass value of $m_N$. For $M_{\tilde N}>2m_N$, $\tilde N\to NN$ decay is possible. For $M_\chi>3m_N$, the decay $\chi\to 3N^ce^+e^-$ is possible.
}\label{fig:massregion}
\end{figure}

For this idea of two DM components to work, we must satisfy the following:
\begin{itemize}
\item The annihilation through Fig.  \ref{fig:chiNRscatt} should be allowed. Namely, $M_\chi+m_N>m_E+m_e$.
\item $\chi$ or $N$ should not decay by the diagram Fig. \ref{fig:chiNRscatt}, and $M_\chi<m_N+m_E+m_e$ should be satisfied for $M_\chi >m_N$. The case $m_N>M_\chi$ turns out to be impossible.
\item The $E$ decay is allowed by the interaction (\ref{eENfromR}): $E_L\to e_R+\tilde N_R$, requiring $m_E>M_{\tilde N}+m_e$.
\item The bino decay $\chi\to 3N^c+e^++e^-$ is forbidden kinematically, $M_\chi<3m_N+2m_e$.
\end{itemize}

By the interaction (\ref{eENfromR}), $E_R^c$ by $\tilde N_R$ propagator ($E_R^c\to e_L^+\tilde N_R^*\to e_L^+N_RN_R$) and $\tilde N_R^*$ directly ($\tilde N_R^*\to 2N_R$ by (\ref{eENfromR})) decay to $2N_R+e_L^+$ and $2N_R$, respectively. For the latter two body decay the total decay rate is
\begin{eqnarray}
\Gamma(\tilde N_R^*)=\frac{|h|^2 M_{\tilde N}\beta_N}{16\pi}\left(1-\frac{2m_N^2}{ M_{\tilde N}^2}
\right)
\label{Ntildedecay}
\end{eqnarray}
where $\beta_N=(1-{m_N^2}/{4 M_{\tilde N}^2})^{1/2}$.
The mass limit  $m_E> 102.5$ GeV for a stable heavy lepton $E$ \cite{PData08} does not apply here. But for our unstable $E$, let us use this bound as a guideline.
For the \LN\ mass we take the usual MSSM estimate of order 100 GeV. Here, we assume that masses of $N$ and $\tilde N_{R,L}$ are relatively small such that the needed kinematics are satisfied. Neglecting the electron mass, in Fig. \ref{fig:massregion} we plot the allowed region in the $M_\chi-m_E$ plane for a specific $m_N$. As far as the decay rate is large enough so that $E$ and $\tilde N$ decayed before 1 second, these decays are not problematic in the nucleosynthesis. Note that given an allowed phase space this is easily satisfied with not too small couplings $f$ and $h$.

If $\chi$ and $N$ each constitutes 50\% of the CDM density, the annihilation diagrams of Fig. \ref{fig:chiNRscatt} account for $\frac12$
of possible encounters of $\chi$ and $N$: $\chi\chi,NN,\chi N, $ and $N\chi$. Among these, the dominant contributions come from $\chi N_R$ and $N_R N_R^c$ annihilation. In principle, the ratio of $\chi$ and $N$ abundance is determined if high energy dynamics is completely known.

Using the interaction (\ref{eENfromR}), the cross section ${d\sigma}(\chi N\to e^+E^-)/{d\Omega}$
of Fig. \ref{fig:chiNRscatt}, in the center of momentum frame with the incident three momentum {\bf p}, is calculated in the small $|\bf p|$ and the large $M_{\tilde E}$ limit as
\begin{eqnarray}
\frac{d\sigma}{d\Omega}&\simeq &\frac{|g'f|^2}{128\pi^2}
\frac{1}{|{\bf p}|}\frac{m_\chi m_N\sqrt{s}\beta_E^4(1+m_E/\sqrt{s})^2}{\left[m_{\tilde e}^2-m_{\chi}^2+m_\chi\sqrt{s}\beta_E^2\right]^2}
\nonumber\\
&&\times\Big\{1+\Big(\frac{2\sqrt{s}\beta_E^2}{m_{\tilde e}^2-m_\chi^2+m_\chi\sqrt{s}\beta_E^2} -\frac{1}{m_\chi}\nonumber\\
&&-\frac{\beta_E^2}{m_N(1+m_E/\sqrt{s})^2}\Big)|{\bf p}|\cos
\theta\Big\} ,\label{crsection3}
\end{eqnarray}
where $\theta$ is the angle between three momenta of $\chi$ and $e^+$, and  $\beta_E^2=1-{m_E^2}/{s}$
with $\sqrt{s}=M_\chi+ m_N$. [We included the $NN^c$ annihilation process also, which however is suppressed for a large $\tilde E_R$ mass.] A similar expression holds for the charge conjugated final states. Then, the velocity averaged cross section is calculated as
\begin{eqnarray}
\langle\sigma v\rangle &\simeq& \frac{|g'f|^2}{32\pi^2}
\frac{m_\chi m_N\beta_E^4(1+m_E/\sqrt{s})^2} {\left[m_{\tilde
e}^2-m_\chi^2+m_\chi\sqrt{s}\beta_E^2\right]^2}
+{\cal O}(v^2) .
\end{eqnarray}

\begin{figure}[!]
\resizebox{0.9\columnwidth}{!}
{\includegraphics{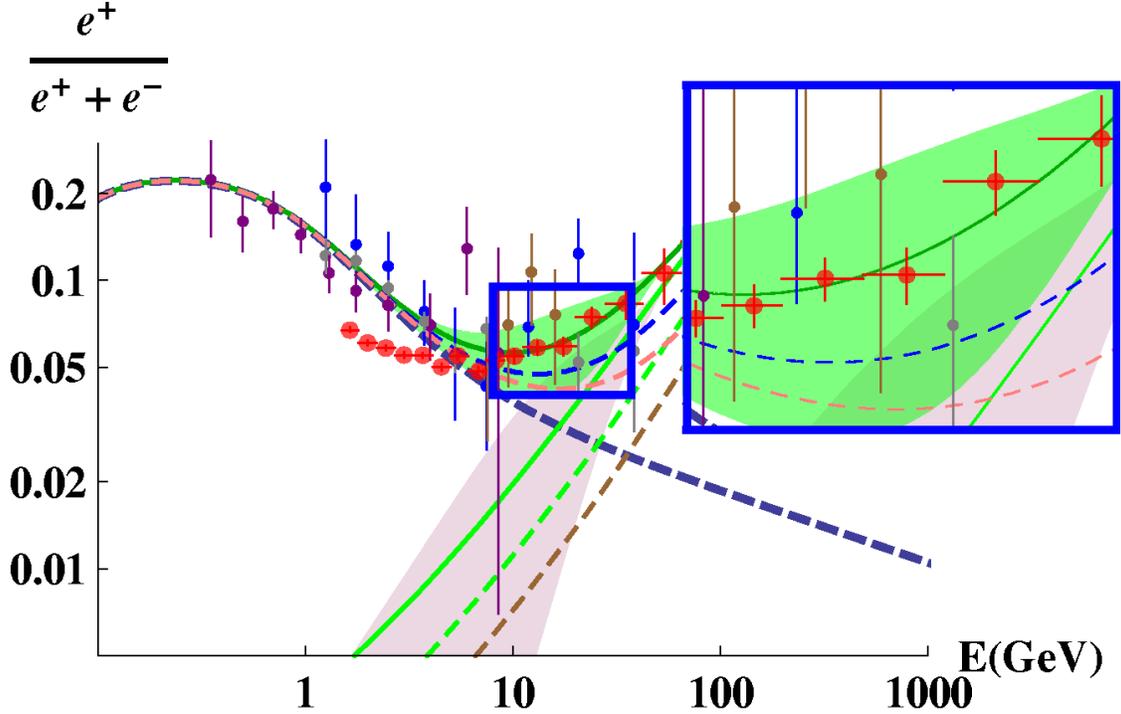}}
\caption{\it
The positron fraction from our model with $M_\chi=200$ GeV, $m_N=80$ GeV, $m_E=200$ GeV, $M_{\tilde E}= 400$ GeV and $M_{\tilde e}=220$ GeV (thick green line) and $B=7$. $M_{\tilde e}$ for 250 GeV (blue dash line) and 280 GeV (brown dash line) are also shown. The pink band is the positron fraction coming from $\chi N$ and $NN$ annihilations and the green band is this positron excess on top of the astrophysical background (the thick darkblue dash line) \cite{Delahaye08,Baltz98}. The width of the band shows the uncertainty from the positron propagation model. The PAMELA data are the red dots \cite{PAMELAexp}, and the various small dots represent the observed positron cosmic ray data \cite{Grimani02,AMS01,CAPRICE94,HEAT95}. }\label{fig:posexcess}
\end{figure}

In Fig. \ref{fig:posexcess}, we present our estimate of the positron excess for typical values of $M_\chi=200$ GeV, $m_N=80$ GeV, $m_E=200$ GeV, $M_{\tilde E}=400$ GeV, and $M_{\tilde e}=$ 220 GeV (thick green line), 250 GeV (blue dash line), 280 GeV (brown dash line). For a good fit, we need a small difference for $M_{\tilde e}-M_\chi$.

To compare with the observations, basically we use the astrophysical background flux given by $\Phi^{bkg}_{e^+}=4.5E^{0.7}/(1+ 650 E^{2.3}+1500 E^{4.2})$ and $\Phi^{bkg}_{e^-}=0.16E^{-1.1}/(1+11E^{0.9}
+3.2E^{2.15})+0.70E^{0.7}/(1+110E^{1.5}
+580E^{4.2})$ \cite{Baltz98,Moskalenko98}.
The deviation of the PAMELA data from this curve at low energy($<10$ GeV) can be explained by the solar modulation effect \cite{PAMELAexp}. The calculation of the positron flux from a given particle physics model is well described in Refs. \cite{Delahaye08,Cirelli08}. The positron flux is given by $\Phi_{e^+}=v_{e^+}\xi/4\pi$, where $v_{e^+}$ is the velocity of the positron and $\xi$ is the positron number density per unit energy, $\xi=dN_{e^+}/dE$. $\xi$ is determined by the diffusion-loss equation using the various cosmic ray data as described in \cite{Delahaye08}.
Under the steady state approximation, the solution of the diffusion-loss equation is given by a semi-exact form
$$
\Phi_{e^+}\hskip -0.05cm=\frac{Bv_{e^+}}{4\pi b(E)}
\hskip -0.05cm\int^{\infty}_{E}\hskip -0.25cm
dE'\sum_{i,j}\langle\sigma v\rangle_{i,j}\frac{\rho^2}{m_i m_j}\frac{dN}{dE'} I(\lambda_D(E,E'))
$$
where $I(\lambda_D)$ is the halo function which has the halo model dependence but is independent from particle physics and $B\geq1$ is a possible boost factor coming from the DM halo substructure \cite{Delahaye08}.

In conclusion, it is likely that the \LN\ of the MSSM cannot explain the high energy positron spectrum of the PAMELA/HEAT data \cite{PAMELAexp,HEAT95}, simply because of the angular momentum constraint. So, we extended the MSSM to keep low energy supersymmetry with additional fields $N$ and $E$ with the $U(1)_R$ symmetry, to obtain two DM particles $\chi$ and $N$. It is the minimal extension toward explaining the PAMELA/HEAT data. In the
$N_{\rm DM}$MSSM, it is shown that a wide range of the parameter space for two DM components is allowed.

\acknowledgments{
This work is supported in part by the Korea Research Foundation, Grant No. KRF-2005-084-C00001. B.K. is also supported by the FPRD of the BK21 program (No. K20732000011-07A0700-01110).}


\end{document}